\documentclass[letterpaper,12pt]{article}
\usepackage{graphicx}
\usepackage{color,amssymb,enumerate,float,amsmath}
\usepackage{tikz}
\usetikzlibrary{matrix}
\usepackage{caption}
\usepackage{subcaption}
\usepackage{}
\usepackage{ifpdf}
\ifpdf
	\usepackage[pdftex,unicode,implicit]{hyperref}

	\hypersetup{
  	pdftitle     = {}, 
  	pdfkeywords  = {},
  	pdfauthor    = {},
  	pdfcreator   = {pdf\LaTeXe\ with package \flqq hyperref\frqq},
  	pdfproducer  = {pdf\LaTeXe\ with package \flqq hyperref\frqq},
  	pdfpagemode  = UseNone,  
  	pdffitwindow = true,  
  	unicode      = true,
  	plainpages   = true,
  	colorlinks   = true,  
  	citecolor    = black,  
  	urlcolor     = blue, 
  	linkcolor    = black
	}

\else

  \usepackage[dvips]{graphicx}
  \usepackage[unicode,implicit]{hyperref}

\fi 

\makeatletter
\@addtoreset{equation}{section}
\makeatother

\begin{document}
	
\thispagestyle{empty}

\begin{center}
{\bf \LARGE Casimir effect of a rough membrane in an aether-like Lorentz-violating scenario }
\vspace*{15mm}

{\large Byron Droguett}$^{1,a}$
{\large and Claudio B\'orquez}$^{2,b}$
\vspace{3ex}

$^1${\it Department of Physics, Universidad de Antofagasta, 1240000 Antofagasta, Chile.}

$^2${\it Facultad de Ingenier\'ia, Arquitectura y Diseño, Universidad San Sebasti\'an, Lago Panguipulli 1390, Puerto Montt, Chile.
}

\vspace{3ex}

$^a${\tt byron.droguett@uantof.com},\quad
$^b${\tt claudio.borquez@uss.cl}\hspace{.5em}

\vspace*{15mm}
{\bf Abstract}
\begin{quotation}{\small\noindent
We explore the Casimir effect of a rough membrane within the framework of theories that break Lorentz symmetry. We consider two constant aether vectors: one timelike and other spacelike, simultaneously. We employ an appropriate change of coordinates such that the membrane assumes a completely flat border and the remaining terms associated with the roughness are considered as part of the potential. Quantum fluctuations are induced by a scalar quantum field subject to Dirichlet boundary conditions. The spectrum is obtained through perturbation theory and regularized using the $\zeta$--function method. We provide an explicit example of a membrane with periodic boundaries. The presence of aether vectors has a significant impact on the dominant term of the Casimir effect, while roughness only affects the secondary terms. Additionally, we examine the finite-temperature case.}

\end{quotation}
\end{center}

\thispagestyle{empty}

\newpage

\section{Introduction}
The Casimir effect is a physical manifestation predicted by quantum field theory resulting from vacuum quantum fluctuations. H. B. Casimir demonstrated that two parallel plates, uncharged and isolated separated by a distance much smaller than their length, experience an attractive force between them due to quantum fluctuation of the electromagnetic field under certain boundaries conditions \cite{Casimir:1948dh}. This phenomenon has been demonstrated with a high degree of accuracy, making it a good experiment to study the properties of vacuum quantum fluctuations \cite{Lamoreaux:1996wh, Bressi:2002fr}. Other geometries have been studied, and in some cases, the force is repulsive, indicating that the Casimir effect depends on the geometry \cite{Boyer:1968uf}. The Casimir effect has been studied in various contexts, and it has been demonstrated that different factors affect the spectrum. The research has revealed that boundary conditions, which can represent some type of material, spacetime topology and temperature, also modified the energy spectrum \cite{Bordag:2001qi, Teo:2011kt, Zhao:2006rr}. The dimensionality of spacetime plays a fundamental role in the Casimir effect. For instance, studies on two-dimensional materials as boundaries, such as the graphene family \cite{Bellucci:2019ybj},  are important for technological advancements in materials science.

Our objective is to study the three-dimensional spacetime Casimir effect in quantum field theories considering the breaking of the Lorentz symmetry. The quantum field theories are based on special relativity, hence they are invariant under Lorentz transformations. This symmetry has been demonstrated both theoretically and experimentally in the low-energy regime \cite{Bordag:2009zz, Milton:2001yy}. Despite this, there is still the possibility that in high-energy regimes this symmetry can be broken \cite{Kostelecky:1988zi}, so some trace of this breaking may remain at low energies \cite{Carroll:2001ws, Anisimov:2001zc, Hewett:2000zp, Bertolami:2003nm, Kostelecky:2002ca, Anchordoqui:2003ij,Alfaro:1999wd,Alfaro:2001rb}. Recently, P. Ho\v rava  proposed a quantum field theory of gravity which is heuristically renormalizable and unitarity, at least in power counting \cite{Horava:2009uw, Blas:2009qj}. The central idea is based on the existence of an anisotropy between space and time in the ultraviolet, causing general diffeomorphisms to be reduced and breaking in the Lorentz symmetry. The Casimir effect has been analyzed within the framework of Ho\v rava gravity theory in a prior study \cite{Borquez:2023cuf}. Extensions to Klein-Gordon  and fermionic field theories have been formulated, known as Ho\v rava-Lifshitz type theories \cite{Ferrari:2010dj, MoralesUlion:2015tve, Muniz:2014dga, daSilva:2019iwn, Erdas:2023wzy, Borquez:2023ajx}. Other research on Lorentz violation has incorporated aether-like vectors into the Lagrangian, setting a preferred direction. \cite{Cruz:2017kfo, deMello:2022tuv, Cruz:2018thz, ADantas:2023wxd, Ferreira:2021xzf, Escobar:2020pes, Erdas:2020ilo}. Furthermore, finite temperature effects in these theories
have been explored \cite{Cruz:2018bqt, Erdas:2021xvv, Cheng:2022mwd}. Due to the high experimental precision of the Casimir effect and the large number of theoretical works on the breaking of Lorentz symmetry, experiments associated with the Casimir effect become good candidates for detecting this breakdown.
 
In this research, we analyze the modifications to the Casimir energy spectrum produced by two contributions: the presence of roughness on the membrane embedded in a three-dimensional spacetime manifold and the introduction of two timelike and spacelike unit orthogonal aether vectors. For a realistic application, we consider roughness as a perturbation from the flat case, such that by performing an appropriate change of coordinates, the membrane assumes a completely flat border and the remaining terms associated with the roughness are considered as part of the potential. Likewise, we consider the presence of two orthogonal unit aether vectors  that are included within low-energy terms in the modified Klein-Gordon Lagrangian. These act on the covariant derivatives of scalar quantum fields, causing the breaking of Lorentz symmetry. To find the spectrum of eigenvalues, we employ perturbation theory  and utilize the $\zeta$–function in the regularization process \cite{Kirsten:2010zp}. Furthermore, we present a specific example of a membrane with periodic border. Additionally, we take into consideration the impact of temperature through the effective action.

This paper is organized as follows. In section 2, we present the problem of the
rough membrane considering two orthogonal constant aether vectors and the solution to the eigenvalue problem through the perturbation theory. In section 3, we apply the regularization method using the $\zeta$–function
and determine the energy and force density in the limit of infinite length. We
present an explicit example of a membrane with periodic border. In section 4, we
consider the effects of temperature on the membrane. Finally, in section 5, we
present our conclusions.



\section{Modified spectral values by rough membrane and aether vectors}
We analyze the low-energy case where the Klein-Gordon scalar field theory is modified by the presence of two parameters $\sigma_{1}$ and $\sigma_{2}$. These parameters define a privileged direction in spacetime, thus leading to the violation of Lorentz symmetry.
The  Lagrangian density  of the modified Klein-Gordon theory by two orthogonal unit aether vectors is given by
\begin{equation}
    \mathcal{L}=
    \frac{1}{2}
    \left(
\nabla_\mu\phi\nabla^\mu\phi
+
    \sigma_1
     u_1^\mu\nabla_\mu
     \phi
     u_1^\nu\nabla_\nu
     \phi
     +\sigma_2
     u_2^\mu\nabla_\mu
     \phi
     u_2^\nu\nabla_\nu
     \phi
    \right)
    \,,
\end{equation}
 where $\sigma_1,\sigma_{2}\ll1$ are dimensionless constants, $u^{\mu}_1$ and $u^{\mu}_2$ are unit  aether vectors. $\nabla_\mu$ is the covariant derivative depending of the spacetime metric.
The equation of motion of the modified Klein-Gordon field is given by\footnote{Where the metric signature has been considered as $(+,-,-)$.}
\begin{equation}
    \left(
    \nabla_\mu\nabla^\mu+\sigma_1
    \left(
    u_1^\mu\nabla_\mu
    \right)^2
    +\sigma_2
    \left(
    u_2^\mu\nabla_\mu
    \right)^2
    \right)\phi
    =0
    \,.
    \label{KG}
\end{equation}
With these modifications on the Klein-Gordon equation, we extend the study done in \cite{Borquez:2023ajx} with the presence of aether vectors \cite{Cruz:2017kfo,deMello:2022tuv}.

Our aim is to investigate the Casimir effect resulting from the quantum fluctuations of a scalar field acting on a rough membrane embedded in a flat three-dimensional spacetime manifold including two orthogonal unit timelike and spacelike  aether vectors, simultaneously.
The membrane is modeled by the following coordinates\footnote{It is possible to configure perturbations at both edges of the membrane so that the contributions can be additive or cancel. In this sense, the conclusions about how the aether and perturbation terms affect the Casimir force remain consistent.}:
\begin{equation}
0\leq x\leq L\,, \qquad 0\leq y\leq a+h(x)\,,
\end{equation}
where $L$ denotes the length of the membrane, $a$ represents its width and $h(x)$ contains all the information about the roughness of the membrane, with the assumption that $h(x)\ll a\ll L$.
Taking these conditions into account, we implement a change of variable on the $y$ coordinate, in such a way that the membrane exhibits flat borders. Then, we consider the new dimensional coordinate
\begin{equation}
\rho = \left(\frac{a}{a+h(x)}\right)y,
\end{equation}
and the new coordinates are defined by
\begin{equation}
      0\leq x\leq L\,, \qquad 0\leq \rho\leq a\,,
\end{equation}
hence we can formulate the spatial metric in the new variables
\begin{equation}
g_{ij}=
\begin{pmatrix}
    1+
    \left(
    \frac{h^{'}\rho}{a}
    \right)^2 
    &
    \left(
    1+\frac{h}{a}
    \right) \frac{h^{'}\rho}{a} 
    \\
   \left(
   1+\frac{h}{a}
   \right) \frac{h^{'}\rho}{a} 
   & 
   \left(
   1+\frac{h}{a}
   \right)^2
   \\
\end{pmatrix}
\,.
\end{equation}
From this metric, we can construct the Laplace-Beltrami operator. The roughness of the membrane is considered perturbatively, hence the Laplace-Beltrami operator can be expanded to second order in perturbations in terms of $h(x)/a$. The remaining terms associated with the roughness are included in a potential term \cite{Borquez:2023ajx}.
Moreover, in order to operate with dimensionless coordinates we implement the following parameterization:
\begin{eqnarray}
\begin{split}
    x & = uL\,, \qquad 0\leq u\leq 1\,,
    \\
    \rho & = va\,,\qquad 0\leq v\leq 1\,.
\end{split}
\end{eqnarray}
Since we consider $L\gg h(x)$, we can discard several terms from the operator in the new coordinates, leading to a helpful simplification. Finally, the Laplace-Beltrami operator has the form
\begin{eqnarray}\label{BelSim}
     - \mathcal{P} \equiv
    \frac{1}{L^2}\partial_u^{2}
    + \frac{1}{a^2}\partial_v^{2}
    - \mathcal{M}(u)\partial_v^2
    \,,
    \label{OP}
    \end{eqnarray} 
where
\begin{eqnarray}
    \mathcal{M}(u)= \left(
     \frac{2\hat{h}}{a^{3}}
    - \frac{3\hat{h}^{2}}{a^4}
    \right),
\end{eqnarray}
with $\hat{h}=h(uL)$. In this new coordinates, we impose the following Dirichlet boundary conditions for the scalar field:
\begin{eqnarray}
\begin{split}
        \phi(u,0)& =\phi(u,1) = 0\,,
        \\
        \phi(0,v) & =\phi(1,v) = 0\,.
\end{split}
\label{Dirichlet}
\end{eqnarray}

The presence of the aether vectors in Eq. (\ref{KG}) modify the general structure of the operator $\mathcal{P}$ in Eq. (\ref{OP}). In this study, we choose temporal $u_{1}=(1,0,0)$ and spatial $u_{2}$ vectors simultaneously, where $u_{2}$ can be: $u_{2}=(0,1,0)$ or $u_{2}=(0,0,1)$, which correspond to parallel and perpendicular vectors to the membrane length, respectively \cite{Cruz:2017kfo,deMello:2022tuv}. The modifications made by the aether vectors define two new spatial operators: $\mathcal{P}_{\parallel}$ and $\mathcal{P}_{\perp}$. To solve this eigenvalue problem associated to spatial operator in the presence of a rough membrane and the aether vectors, we use the perturbation theory.
For the zeroth order in perturbations, the parallel and perpendicular spatial operators are respectively:
\begin{eqnarray}
        \left(-\frac{1}{L^2}\left(1-\sigma_2\right)\partial_u^{2}
    -\frac{1}{a^2}\partial_v^2\right)\phi^{(0)}=(1+\sigma_1)\lambda^{(0)}_{\parallel}\phi^{(0)}
    \,,
    \label{SpatialOperatorparal}
    \end{eqnarray}
    \begin{eqnarray} 
        \left(-\frac{1}{L^2}\partial_u^{2}
    -\frac{1}{a^2}\left(1-\sigma_2\right)\partial_v^2\right)\phi^{(0)}=(1+\sigma_1)\lambda^{(0)}_{\perp}\phi^{(0)}
    \,.
     \label{SpatialOperatorPerp}
    \end{eqnarray}
The constant $\sigma_1$ reflects the existence of a preferred temporal direction and it is global factor. For both cases the normalized solution is
\begin{eqnarray}
    \phi_{n,m}^{(0)}(u,v)=2\sin\left(n\pi v\right)\sin\left(m\pi u\right)\,,
\end{eqnarray}
and by considering the Dirichlet boundary conditions in (\ref{Dirichlet}), the eigenvalues correspond to
\begin{equation}
\begin{split}
        \lambda_{n,m,\parallel}^{(0)}
        =&
        \frac{1}{1+\sigma_1}\left[
        \left(1-\sigma_2\right)^2\left(\frac{m\pi}{L}\right)^2
        +
        \left(\frac{n\pi}{a}\right)^2
        \right]
        \,,
        \\
        \lambda_{n,m,\perp}^{(0)}
        =&
        \frac{1}{1+\sigma_1}\left[
        \left(\frac{m\pi}{L}\right)^2
        +
        \left(1-\sigma_2\right)^2\left(\frac{n\pi}{a}\right)^2
        \right]
        \,.
\end{split}
\label{autova0}
\end{equation} 
In general, the following perturbation orders can be calculated, and these higher-order terms contribute significantly. Although our method can be applied to any type of roughness, from this point on, we will focus exclusively on periodic functions for $\hat{h}$. With this assumption, the only perturbation term contributing to the eigenvalues comes from the first-order perturbations, that is:
\begin{eqnarray}\label{per1}
        -\frac{(n\pi)^2}{1+\sigma_1}
        \int_0^1\,du\mathcal{M}(u)
        \,.
    \end{eqnarray}
    
By including the aether vectors in the general operator (\ref{KG}), the function $\mathcal{M}$ is modified, thus, for the parallel and perpendicular cases, the following is established
    \begin{equation}
       \mathcal{M}_{\parallel}
       =
       \mathcal{M}
    \,,
    \quad
       \mathcal{M}_{\perp}
       =
       (1-\sigma_2)\mathcal{M} 
    \,.
    \end{equation}
The coupling constant $\sigma_2$ indicates that the aether term contributes to the first order in perturbation theory only when the aether vector is perpendicular to the length of  the membrane.
Therefore, from Eqs. (\ref{KG}), (\ref{OP}), (\ref{autova0}) and (\ref{per1}), the total eigenvalues are given by      
    \begin{eqnarray}
        \lambda_{n,m,\parallel}=
        \frac{1}{1+\sigma_1}\left[
        \left(
        \frac{n\pi}{a}
        \right)^2
        +(1-\sigma_2)^2\left(
        \frac{m\pi}{L}
        \right)^2
        -(n\pi)^2\int_0^1\,du\mathcal{M}_{\parallel}\right]
        \,,
        \nonumber
        \\
        &&
        \label{totaleingenvalparal}
    \end{eqnarray}
     \begin{eqnarray}
        \lambda_{n,m,\perp}=
        \frac{1}{1+\sigma_1}\left[
         (1-\sigma_2)^2\left(
       \frac{n\pi}{a}
        \right)^2
        +\left(
        \frac{m\pi}{L}
        \right)^2
        -(n\pi)^2\int_0^1\,du\mathcal{M}_{\perp}\right]
           \,.
        \nonumber
        \\
        &&
        \label{totaleingenvalperp}
    \end{eqnarray}

\section{Casimir effect: $\zeta$--function regularization}
We will address the regularization of the spectrum generated by both parallel and perpendicular spatial operators using the $\zeta$--function method, which are respectively given by 
\begin{eqnarray}
\label{ZetaP//}
    \zeta_{\mathcal{P}_{\parallel}}(s)
    =(1+\sigma_1)^s\sum_{n,m\in\mathbb{N}}\left[
    \left(
 \frac{n\pi}{a}
        \right)^2
        +(1-\sigma_2)^2\left(
        \frac{m\pi}{L}
        \right)^2
        -(n\pi)^2\int_0^1\,du\mathcal{M}_{\parallel}
        \right]^{-s}\,,
\end{eqnarray}
\begin{eqnarray}
\label{ZetaPperp}
    \zeta_{\mathcal{P}_{\perp}}(s)
    =
    (1+\sigma_1)^s
    \sum_{n,m\in\mathbb{N}}\left[
    (1-\sigma_2)^2\left(
        \frac{n\pi}{a}
        \right)^2
        +\left(
        \frac{m\pi}{L}
        \right)^2
        -(n\pi)^2\int_0^1\,du\mathcal{M}_{\perp}
        \right]^{-s}\,.
\end{eqnarray}
These $\zeta$--functions have the structure of the Epstein $\zeta$--function and can be represented in the general integral form 
\begin{eqnarray}
    \zeta_{\mathcal{P}_{\parallel,\perp}}(s)
    =
    \frac{(1+\sigma_1)^s}{\Gamma(s)}
    \int_0^\infty\,dt\, t^{s-1}
    \sum_{n,m\in\mathbb{N}}\exp\left[-t\left(r_1n^2+r_2m^2\right)\right],
    \label{zetaPP}
\end{eqnarray}
where the $r_1$ and $r_2$ functions for the parallel and perpendicular cases are
\begin{eqnarray}\label{R12paral}
    r_{1,\parallel}
    =
    \pi^2\left(
    \frac{1}{a^2}-\int_0^1\,du\mathcal{M_{\parallel}}
    \right)
    \,,
    \qquad
    r_{2,\parallel} =
    \left(1-\sigma_2\right)^2\left(
    \frac{\pi}{L}
    \right)^2\,,
\end{eqnarray}
\begin{eqnarray}\label{R12perp}
    r_{1,\perp}
    =
    \pi^2\left(
    \frac{(1-\sigma_2)^2}{a^2}-\int_0^1\,du\mathcal{M}_{\perp}
    \right)
    \,,
    \qquad
    r_{2,\perp} =
    \left(
    \frac{\pi}{L}
    \right)^2\,.
\end{eqnarray}
To continue with the regularization we must perform an analytical extension on the $\zeta$--function in (\ref{zetaPP}), through the Poisson summation technique \cite{Kirsten:2010zp}
\begin{eqnarray}
    \zeta_{\mathcal{P}_{\parallel,\perp}}(s)
    &=&
    \frac{(1+\sigma_1)^s}{4\Gamma(s)}\left\{\int_0^\infty\,dt\,t^{s-1}
    + \frac{2\pi}{\sqrt{r_{1}r_{2}}}\int_0^\infty\,dt\,t^{s-2}
    \left[\frac{1}{2} + \sum_{n=1}^{\infty}e^{\frac{-\pi^{2}n^{2}}{tr_{1}}} + \sum_{m=1}^{\infty}e^{\frac{-\pi^{2}m^{2}}{tr_{2}}}\right.\right.
    \nonumber
    \\
    &&
    \left.\left.
    + 2\sum_{n,m=1}^{\infty}e^{-\frac{\pi^{2}}{t}\left(\frac{n^{2}}{r_{1}}+\frac{m^{2}}{r_{2}}\right)}\right]
    - 2\sqrt{\pi}\int_0^\infty\,dt\,t^{s-3/2}\left[\frac{1}{2}\left(\frac{\sqrt{r_1}+\sqrt{r_2}}{\sqrt{r_1r_2}}\right)\right.\right.
    \nonumber
    \\
    &&
    \left.\left.
    + \frac{1}{\sqrt{r_{1}}}\sum_{n=1}^{\infty}e^{\frac{-\pi^{2}n^{2}}{tr_{1}}}
    + \frac{1}{\sqrt{r_{2}}}\sum_{m=1}^{\infty}e^{\frac{-\pi^{2}m^{2}}{tr_{2}}}
    \right]\right\}
    \,.
    \label{zetaPs}
\end{eqnarray}
The Casimir energy is calculated  by evaluating the $\zeta$--function at $s=-1/2$. The consequences of evaluating this condition is that the integrals generate divergent terms, which we must remove in order to obtain a finite result.
In this way, the Casimir energy is expressed as $E_{C}=\frac{1}{2}\zeta_{\mathcal{P}_{\parallel,\perp}}(-1/2)$, hence we have for both cases 

\begin{eqnarray}
    E_{C_{\parallel,\perp}}
    &=&
    -\frac{(1+\sigma_1)^{-\frac{1}{2}}}{8\pi^2}\left\{\frac{1}{2}\zeta_{R}(3)\left(\frac{r_{1}}{\sqrt{r_{2}}}+\frac{r_{2}}{\sqrt{r_{1}}}\right)
    + r_{1}r_{2}\sum_{n,m=1}^{\infty}\left(r_{2}n^{2} + r_{1}m^{2}\right)^{-3/2}
    \right.
    \nonumber
    \\
    &&
    \left.- \zeta_{R}(2)\left(r_{1}^{1/2} + r_{2}^{1/2}\right)\right\}
    \,.
\end{eqnarray}
The energy density per unit length is considerably reduced by tending $L$ to infinity, thus, we have 
\begin{eqnarray}
    \mathcal{E}_{C_{\parallel}}
    &=&
    - \frac{(1+\sigma_{1})^{-1/2}}{1-\sigma_2}\frac{\zeta_{R}(3)}{16\pi}\lim_{L\rightarrow\infty}
    \left[
    \frac{1}{a^2}
    - \int_0^1\,du\mathcal{M}_{\parallel} 
    \right]
    \,,
\end{eqnarray}
\begin{eqnarray}
    \mathcal{E}_{C_{\perp}}
    &=&
    - (1+\sigma_{1})^{-1/2}\frac{\zeta_{R}(3)}{16\pi}\lim_{L\rightarrow\infty}\left[
    \frac{(1-\sigma_2)^{2}}{a^2}
    - \int_0^1\,du\mathcal{M}_{\perp} 
    \right]
    \,.
\end{eqnarray}
Consequently, we obtain the Casimir force density by deriving the energy density with respect to the width $a$ of the membrane
\begin{eqnarray}
    \mathcal{F}_{C_{\parallel}}
    &=&
    - \frac{(1+\sigma_{1})^{-1/2}}{1-\sigma_2}\frac{\zeta_{R}(3)}{8\pi}\lim_{L\rightarrow\infty}\left[\frac{1}{a^{3}}-\frac{3}{a^{4}}\int_0^{1}\hat{h}du + \frac{6}{a^{5}}\int_0^{1}\hat{h}^{2}du\right]\,,
    \nonumber\\
    \label{FP1}
\end{eqnarray}
\begin{eqnarray}
    \mathcal{F}_{C_{\perp}}
    &=&
    - \frac{1-\sigma_{2}}{(1+\sigma_{1})^{1/2}}\frac{\zeta_{R}(3)}{8\pi}\lim_{L\rightarrow\infty}\left[\frac{(1-\sigma_{2})}{a^{3}}-\frac{3}{a^{4}}\int_0^{1}\hat{h}du + \frac{6}{a^{5}}\int_0^{1}\hat{h}^{2}du\right]\,.
    \nonumber\\
    \label{FPe1}
\end{eqnarray}

To illustrate these results we will proceed to calculate the Casimir effect for a specific case of a membrane with periodic roughness. Considering the original coordinates, we choose the following periodic function
\begin{eqnarray}
    h(x)=
    \epsilon\cos\left(\frac{2\pi}{L}  x + \theta \right)\,,
    \label{hcos}
\end{eqnarray}
where $\theta$ is an arbitrary phase and $\epsilon$ correspond to a small parameter with length unit that represents the perturbative nature of $h$. The coefficients $\sigma_{1,2}$ associated with the vectors that break the Lorentz symmetry are treated perturbatively, therefore, we must expand around these terms. \textcolor{red}{}

\begin{figure}[H]
    \begin{minipage}[b]{0.5\linewidth}
    \centering
    \includegraphics[width=.9\linewidth]{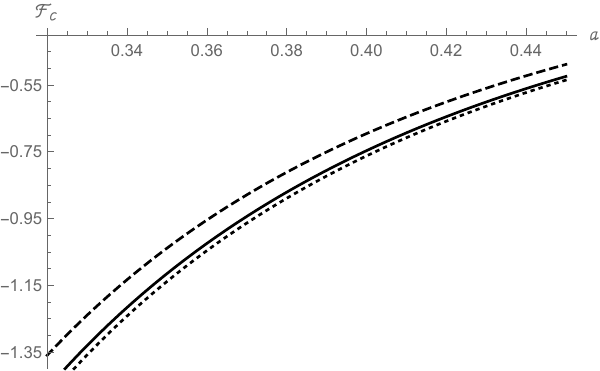} 
    \caption*{(a)}
    \vspace{4ex}
  \end{minipage}
    \begin{minipage}[b]{0.5\linewidth}
    \centering
    \includegraphics[width=.9\linewidth]{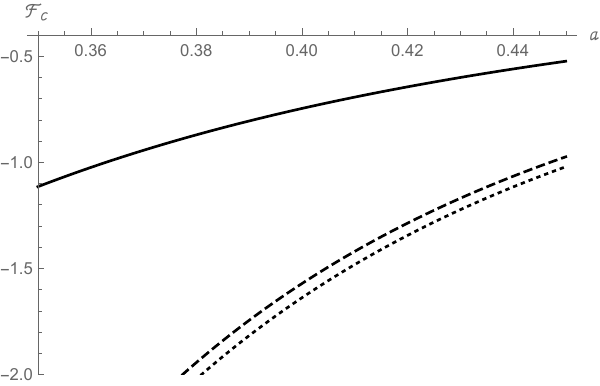}
    \caption*{(b)}
    \vspace{4ex}
  \end{minipage}
  \caption{Both graphs represent the Casimir force density versus separation distance $a$. In figure (a), the effect of Lorentz breaking without considering roughness is shown. The dashed and dotted curves represent the perpendicular and parallel forces, respectively, considering $\sigma_1=0.02$ and $\sigma_2=0.03$. In figure (b), we consider the roughness of the membrane $\epsilon=0.25$ using the same values for $\sigma_{1,2}$ as in figure (a). In both figures the solid curves represents the case $\sigma_1=\sigma_2=0$ and without roughness.}
  \label{fig1}
  \end{figure}

The Casimir force densities are given by
\begin{eqnarray}
    \mathcal{F}_{C_{\parallel}}
    =
    - \frac{1}{8}\frac {\zeta_{R}(3) }{\pi{a}^{3} }
    \left(1 
    - \frac{1}{2}\sigma_{1}
    + \sigma_{2}
    + \frac{3}{8}\sigma_{1}^{2}
    + \sigma_{2}^{2}
    \right)
    - \frac{3}{8}\frac {\zeta_{R}(3) \epsilon^{2}}{\pi{a}^{5}}
    \,,
\end{eqnarray}
\begin{eqnarray}
\mathcal{F}_{C_{\perp}}
    =
    - \frac{1}{8}\frac {\zeta_{R}(3) }{\pi{a}^{3} }
    \left( 1 - \frac{1}{2}\sigma_{1}
    - 2\sigma_{2}
    + \frac{3}{8}\sigma_{1}^{2}
    + \sigma_{1}\sigma_{2}
    + \sigma_{2}^{2}
    \right)
    - \frac{3}{8}\frac{\zeta_{R}(3)\epsilon^{2}}{\pi a^{5}}
    \,.
\end{eqnarray}
In both scenarios, the effects generated by temporal and spatial aether vectors directly impact the primary term $a^{-3}$, making their effects significant. In contrast, roughness only influences the secondary term $a^{-5}$ (see Ref. \cite{Borquez:2023ajx}), hence the effects of the perturbative coefficients $\sigma_{1,2}$ arise independently of $\epsilon$. The magnitude of the parallel force density is always greater than the magnitude of the perpendicular force density by a proportional factor $\sigma_2(3-\sigma_1)$ (see Figure \ref{fig1}). We can see that the equality between the magnitudes of the forces occurs when $\sigma_{2}=0$ or $\sigma_{1}=3$, however, the latter must be discarded because the coefficients have a perturbative nature, that is, $\sigma_{1}\ll 1$.


\section{Casimir effect at finite temperature}

To investigate the effects of temperature on the energy spectrum, we must employ the path integral, where temperature is represented by the imaginary part of time. The path integral for the scalar quantum field is given by
\begin{equation}
    Z=
    \int\mathcal{D}\phi\exp{(S(\phi))}\,.
\end{equation}
From the path integral we can obtain the effective action associated to some differential operator $\mathcal{O}$. The effective action for the parallel and perpendicular spatial aether vectors are given by
\begin{equation}
    \Gamma=
    \frac{1}{2}\ln\det[
    (
    -\partial^{2}_{\tau}
    +\mathcal{P}_{\parallel,\perp})/\mu
    ]=
    \frac{1}{2}\ln\det
    [
    \mathcal{O}_{\parallel,\perp}/\mu
    ]\,,
\end{equation}
where $\tau\in\mathbb{C}$ and $\mu$ is an arbitrary parameter with  mass dimension, introduced to render
the $\zeta$--function dimensionless. Eventually, the $\zeta$--function will be independent of this parameter $\mu$, hence we set $\mu=1$ for simplicity. Therefore, the Casimir energy is defined by
\begin{eqnarray}
   E_{C}
   =
   \frac{\partial}{\partial\xi}\Gamma
   =
   \left.
   -\frac{1}{2}\frac{\partial}{\partial\xi}\left(\frac{d}{ds}\zeta_{\mathcal{O}_{\parallel,\perp}}(s)
   \right)
   \right|_{s=0}
   \,,
   \label{EcTFormula}
\end{eqnarray}
where $\xi=1/T$ is the inverse of the  temperature.

The eigenvalue problem associated to the operator $\mathcal{O}_{\parallel,\perp}$ is expressed by
\begin{equation}
    \left(
    -\partial^{2}_{\tau}
    +\mathcal{P}_{\parallel,\perp}
    \right)\phi
    =\omega\phi
    \,.
\end{equation}
The solution we propose for the  scalar quantum field is
\begin{eqnarray}
   \phi_{k,n,m}(\tau,x^i)=
   \frac{1}{\xi}e^{\frac{2\pi i k}{\xi}\tau}\varphi_{n,m}(x^i)
   \,,
\end{eqnarray}
where the eigenvalues from the time derivative are defined by periodic border, and correspond to $\omega_{k}=\frac{2\pi k}{\xi}$, with $k\in\mathbb{Z}$. Then, the total eigenvalues associated with the operator $\mathcal{O}_{\parallel,\perp}$ are given by
\begin{eqnarray}
   \omega_{k,n,m}
   =
   \left(1+\sigma_{1}\right)\left(
   \frac{2\pi k}{\xi}\right)^{2}
   + 
   \left(r_1n^2
        +r_2m^2
        \nonumber
        \right)\,,
\end{eqnarray}
 where $r_1$ and $r_2$ are defined by (\ref{R12paral})--(\ref{R12perp}). We use the integral representation of the $\zeta$--function to rewrite the spectral function as
\begin{eqnarray}\label{ZetaSum}
   \zeta_{\mathcal{O}_{\parallel,\perp}}(s)
   =
   \frac{1}{\Gamma
   \left(
   s
   \right)}
   \int_{0}^{\infty}dt\,
   t^{s-1}
   \sum_{k=-\infty}^{\infty}
   \sum_{n,m=1}^{\infty}
   \exp\left\lbrace
   -t\left[
   \left(
   1+\sigma_1\right)\left(
   \frac{2\pi k}{\xi}
   \right)^{2}
   +
   \left(r_1n^2+
        r_2m^2
        \right)\right]\right\}\,.
   \nonumber\\
\end{eqnarray}
As was done in the case of zero temperature, it is possible to use Poisson resummation in (\ref{ZetaSum}),
\begin{eqnarray}
 \zeta_{\mathcal{O}_{\parallel,\perp}}(s)
 &=&
 \frac{\xi}{\sqrt{4\pi}}
 \frac{\Gamma\left(
 s-1/2
 \right)}
 {\Gamma\left(s\right)}
\zeta_{\mathcal{P}_{\parallel,\perp}}(s-1/2)
 \nonumber\\
   &&
   +\frac{\xi}{\sqrt{\pi}\sqrt{1+\sigma_1}\Gamma{(s)}}
   \sum_{k,n,m=1}^\infty
   \int_0^\infty\,dt
   \,t^{s-3/2}
   \exp
   \left[
   -\frac{\xi^2 k^2}{4(1+\sigma_1)t}
   -\left(r_1n^2
        +r_2m^2
        \nonumber
        \right)t\right]\,.
        \nonumber\\
\end{eqnarray}
The coupling constant $\sqrt{1+\sigma_{1}}$  is absorbed in $\zeta_{\mathcal{P}_{\parallel,\perp}}$, thus  it is the same as in Eq. (\ref{zetaPs}).
The energy is obtained by taking the derivative of the $\zeta$--function with respect to $s$ and evaluating it at $s=0$. Then, by integrating the exponential function with respect to $t$, we obtain
\begin{eqnarray}
     \zeta_{\mathcal{O}_{\parallel,\perp}}'(0)
    = 
    - \xi\zeta_{\mathcal{P}_{\parallel,\perp}}(-1/2)
    +2\sum_{k,n,m=1}^\infty
   \frac{1}{k}\exp\left[ 
   -\xi k\left(
   r_1n^2+r_2m^2
   \right)^{1/2}/\sqrt{1+\sigma_1}
   \right]\,.
\end{eqnarray}
The sum over $k$ can be explicitly performed using a geometric series, hence the Casimir energy is
\begin{eqnarray}
    E_{C_{\parallel,\perp}}
    =
    \frac{1}{2}\zeta_{\mathcal{P}_{\parallel,\perp}}(-1/2)
   + \frac{1}{\sqrt{1+\sigma_1}}\sum_{n,m=1}^{\infty} 
   \left[
   \frac{\left(
   r_1n^2
   +r_2m^2
   \right)^{1/2}}{\exp\left(
   \xi \left(
   r_1n^2
   +r_2m^2
   \right)^{1/2}/\sqrt{1+\sigma_1}
   \right)-1}
   \right]
   \,.
   \nonumber
   \\
\end{eqnarray}
In this result, it is  crucial to remember that we need to consider the finite terms of $\zeta_{\mathcal{P}_{\parallel,\perp}}$ when we evaluated at $s=-1/2$ (see Eq. (\ref{zetaPs})).

\begin{figure}[H]
    \begin{minipage}[b]{0.47\linewidth}
    \centering
    \includegraphics[width=.93\linewidth]{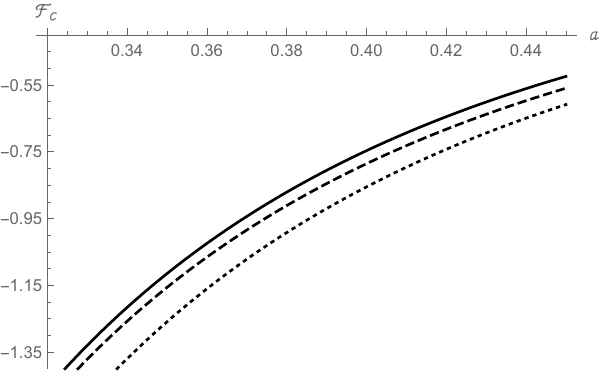} 
    \caption*{(a)}
    \vspace{4ex}
  \end{minipage}
    \begin{minipage}[b]{0.47\linewidth}
    \centering
    \includegraphics[width=.93\linewidth]{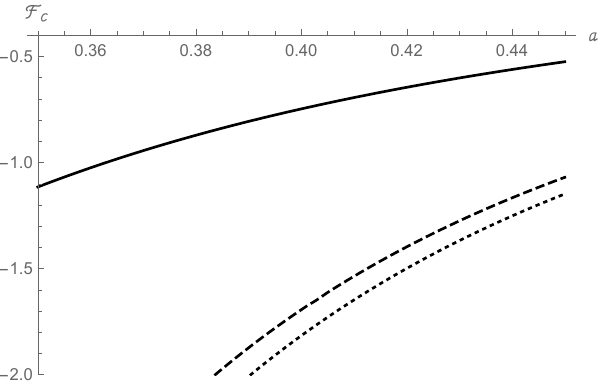}
    \caption*{(b)}
    \vspace{4ex}
  \end{minipage}
  \caption{Both graphs depict the Casimir force density versus separation distance $a$ while taking temperature into account. In Figure (a), the impact of Lorentz symmetry breaking is illustrated without considering roughness. The solid curve represents the scenario where $\sigma_1=\sigma_2=0$ and $T=0$, indicating the absence of Lorentz symmetry breaking. The dashed and dotted lines respectively represent the perpendicular and parallel forces, considering $\sigma_1=0.02$ and $\sigma_2=0.03$, both at the temperature $T=1000$. In Figure (b), we incorporate the roughness $\epsilon=0.25$ of the membrane using the same parameters as in Figure (a).}
  \label{fig2}
  \end{figure}

\begin{figure}[H]
   \begin{minipage}[b]{0.47\linewidth}
    \centering
    \includegraphics[width=.93\linewidth]{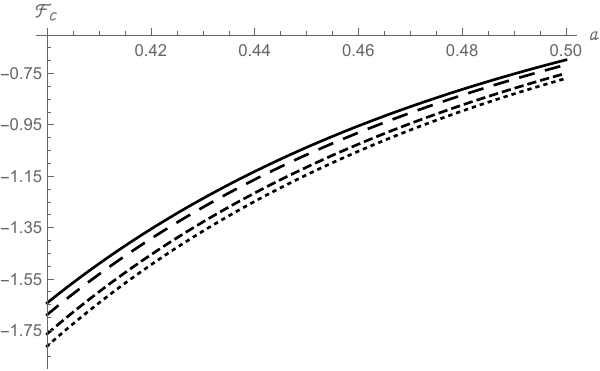}
    \caption*{(a)}
    \vspace{4ex}
  \end{minipage}
    \begin{minipage}[b]{0.47\linewidth}
    \centering
    \includegraphics[width=0.93\linewidth]{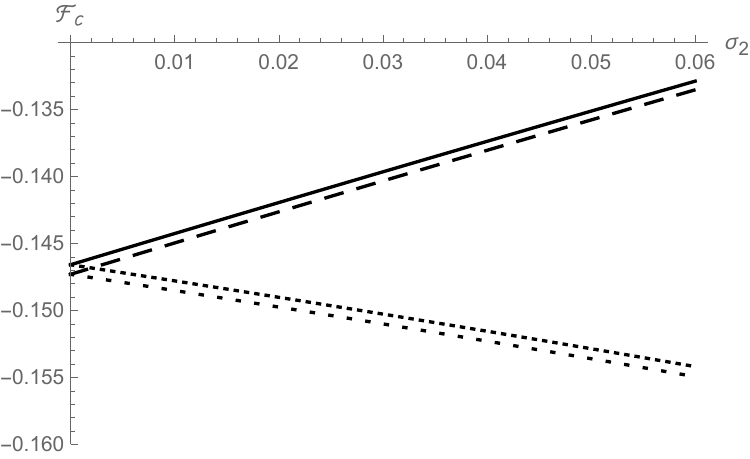}
    \caption*{(b)}
    \vspace{4ex}
  \end{minipage}
  \caption{ The Figure (a) shows the Casimir force density versus separation distance $a$, considering roughness $\epsilon=0.25$, $\sigma_{1}=0.02$, $\sigma_{2}=0.03$ and different temperatures.  The solid and long-dashed curves represent the perpendicular force for $T=10$ and $T=100$, respectively. The dashed and dotted curves show the parallel force for $T=10$ and $T=100$. In the Figure (b), we evaluate the Casimir force density in function of $\sigma_2$ for both  parallel  and perpendicular forces considering temperature $T=100$ and roughness $\epsilon=0.25$. The solid and long-dashed curves represent the perpendicular force  for the cases $\sigma_1=0.03$ and $\sigma_1=0.02$. The dotted and long-dotted curves are the parallel forces  for the cases $\sigma_1=0.03$ and $\sigma_1=0.02$, respectively.}
  \label{fig3}
  \end{figure}

The Casimir force is obtained by deriving the energy with respect to the width $a$ of the membrane. To  simultaneously analyze the effects of roughness, aether vectors and temperature, we will proceed with numerical calculations. In Figure \ref{fig2}, both graphs represent the Casimir force density with respect to the separation of the membrane. In Figure \ref{fig2} (a), the effect of temperature, without considering roughness, increases both parallel and perpendicular force density. In this case the parallel force magnitude always greater than the perpendicular force magnitude. In Figure \ref{fig2} (b), we see that roughness affects the same way, increasing the intensity of the forces. In Figure \ref{fig3} (a) we can see several behaviors of the force density for different values of temperature. The parallel force remains greater than the perpendicular force even when considering the temperature. In Figure \ref{fig3} (b), it  illustrates how the force varies in relation to the values of the constant $\sigma_2$ associated with the spatial aether vectors, for two different values of $\sigma_{1}$ considering temperature and roughness. The solid  and  dotted curves represent the perpendicular and parallel force density respectively, for the case $\sigma_1=0.3$. Both curves coincide when the value of $\sigma_2$ is zero. As the value of $\sigma_2$ increases, the perpendicular force  density decreases, unlike the parallel force, which increases in absolute value. When the value of the constant $\sigma_1$ decreases, both force densities increase in absolute magnitude.

\section{Conclusions}
In this research, we investigate the influences of roughness, aether vectors, and temperature. We analyze the scenario of a rough membrane inserted into a three-dimensional spacetime manifold, where Lorentz symmetry is broken by the presence of both temporal and spatial unit aether vectors. Vacuum quantum fluctuations are induced by a scalar quantum field whose Lagrangian density is formulated within the framework of theories with Lorentz symmetry breaking. To address roughness, we perform a change of variable such that the membrane becomes completely flat border, and the remaining terms associated with perturbative roughness are treated as a potential. The Lagrangian density of the modified Klein-Gordon field incorporates two types of aether vectors: one temporal and one spatial. The considered spatial vector has two possible orientations: parallel or perpendicular to the length of the membrane. We regularize the obtained spectrum using the $\zeta$--function method. Then, we present a specific case of a membrane with periodic border that satisfies Dirichlet boundary conditions. With this we derive two spectra, one associated with the parallel spatial vector and another perpendicular to the length of the membrane. Finally, considering all the previous geometric aspects, we introduce the temperature through the effective action.

The coefficient associated with the timelike vector emerges as a global factor of the standard Casimir effect. On the other hand, for the spacelike vector, the modification on the force and energy is more delicate, because the components parallel and perpendicular have different behavior with respect to the coefficient $\sigma_{2}$. In the perpendicular force density we can see that $\sigma_{2}$ directly affects the primary term, as shown in Eq. (\ref{FPe1}). When we consider the coefficients $\sigma_{1,2}$ perturbatively, our results indicate that the parallel force density is greater than the perpendicular force density. Modifications that break Lorentz symmetry through aether vectors affect the primary term of Casimir energy and force, while the effects of roughness remain isolated, manifesting only in the second order in a secondary term. Thus, the effects of perturbative coefficients $\sigma_{1,2}$ arise independently of $\epsilon$. 
Consequently, the effects induced by the aether vectors exert a significant influence on the Casimir effect, both in its parallel and perpendicular components, and in different ways. We can appreciate this by increasing the value of the parameter $\sigma_2$, the perpendicular force density decreases compared to the parallel case. For both types of force, reducing the value of $\sigma_1$ increases both magnitudes. Considering the effects of temperature in all cases, we observe an increase in the magnitude of the force density.


\end{document}